\documentstyle[preprint,prd,tighten,eqsecnum,aps]{revtex}
\begin{document}
\preprint{\vbox{\hbox{McGill/93--8}\hbox{TPI--MINN--93/19--T}}}
\draft
\title{\bf{Lepton pairs from thermal mesons}}

\author{Charles Gale~\cite{CG}}
\address{
Physics Department, McGill University\\
3600 University St., Montr\'eal, QC,
H3A 2T8, Canada}

\author{Peter Lichard~\cite{PL}}
\address{
Theoretical Physics Institute, University of Minnesota \\
Minneapolis, MN 55455, USA \\
and \\
Department of Theoretical Physics, Faculty of Mathematics and Physics\\
Comenius University, CS-842 15 Bratislava, Slovakia~\cite{pa}}

\maketitle

\begin{abstract}
We study the net dielectron production rates from an ensemble of thermal
mesons,
using an effective Lagrangian to model their interaction. The coupling between
the electromagnetic and the hadronic sectors is done through the vector meson
dominance
approach. For the
first time, a complete set of light mesons is considered. We include
contributions from decays of the type V~(PS)~$\rightarrow$~PS~(V)~+~$e^+~e^-$,
where V is a vector meson and PS is a pseudoscalar, as well as those from
binary reactions PS~+~PS, V~+~V, and V~+~PS~$\rightarrow~e^+e^-$. Direct decays
of the type V~$\rightarrow~e^+ e^-$ are included and shown to be important.
We find that the dielectron invariant mass spectrum
naturally divides in distinct regions: in the low mass domain
the decays from vector and
pseudoscalar mesons form the dominant
contribution. The pion--pion annihilation and direct decays then pick up and
form the leading
signal in an invariant mass region that includes the $\rho - \omega$ complex
and extends up to the
$\phi$. Above invariant mass
$M\ \approx$~1~GeV other
two-body reactions take over as the prominent mechanisms for lepton pair
generation. These facts will have quantitative bearing on the
eventual identification of the quark--gluon plasma.

\end{abstract}

\narrowtext

\section{Introduction}
\label{sec:intro}

One of the ultimate goals of high energy heavy ion physics
is the formation and observation of a quark--gluon plasma (QGP)
as predicted by QCD. A vigorous experimental program
is under way and it is
fair to say that this area is one of the most active fields of contemporary
subatomic physics. The creation of such a novel state of matter
represents a considerable challenge, both in its experimental realization
and also in the theoretical interpretation of the experimental results. The
lifetimes involved are of the order
of $\approx$ 10 fm/c and the detailed dynamics
of the collision process may furthermore play an important role, complicating
the extraction of a clear signal. Nevertheless, much progress has been made
both in theory and in experiment, and we may say
that even in the
absence of a genuine QGP the study of hot and dense hadronic systems is
still a fascinating subject from which a great deal can be learnt.

For a while now,
electromagnetic signals have been known as ideal probes of strongly
interacting matter at high temperatures and densities \cite{rev}. This owes to
the fact that once they are produced, they will travel relatively unscathed
from their point of origin to the detector. Since production rates are rapidly
increasing functions of temperature and density, these electromagnetic signals
provide valuable information on the hot and dense phases of the reaction. It is
hoped that, because of these facts, those signals should constitute precious
aids in the process of analyzing the behaviour of hot quark--gluon
matter \cite{rus92}. As with any possible experimental signature of the
QGP, a great deal of
care must go into the calculation of a corresponding ``purely hadronic''
signal, that is a contribution to the same experimental observables
from sources other
than the deconfined, chiral--symmetric phase. As far as the quark--gluon plasma
is concerned, one may refer to these sources as the ``background''.

In this paper, we are concerned with the thermal rate of dielectron emission
only but our treatment
is completely general. The source is a hot environment of several meson
species: for the first
time, we use a rather complete set of mesons, rather than restricting ourselves
to the usual pion gas approximation. The equilibrium assumptions inherent to
the approaches similar to the one been used here have to be carried to their
logical conclusion: in such scenarios, once the temperature has been set one
can clearly calculate the population of species present. These mesons can then
interact among themselves, or even decay, to produce lepton pairs in the
final state. It is important to realize that we deliberately make no attempt
here to connect
with experiment because our calculation is rather meant to answer a well
defined theoretical question: what is the electromagnetic emissivity (in the
dilepton
channel) of a hot hadron gas? To answer this question, we shall
proceed along the lines of a similar
calculation for photon rates \cite{kap91}.

We estimate the rates of producing lepton
pairs using relativistic kinetic theory. The mesonic interactions are modelled
with an effective Lagrangian and the coupling of radiation to hadronic matter
is done in the vector meson dominance (VMD) approach. The
values
of the coupling
constants involved are adjusted so that the experimentally measured radiative
decay widths are reproduced. We describe the details of our model in section
\ref{sec:model}. We give results in section \ref{sec:results} and finally we
end with a discussion in section \ref{sec:conclusion}.

\section{The model}
\label{sec:model}

Our starting point is an ensemble of mesons in thermal equilibrium.
We consider the lightest and thus most abundant strange and non--strange mesons
together with their main interaction channels. This means we shall include:
$\pi, \eta, \rho,
\omega, \eta^\prime, \phi, K$ and $K^*$. The charge states are not labelled but
all of them are present. This collection can be further divided in two
categories:
pseudoscalar (PS) and  vector (V) particles.
{}From this
hot meson gas, how do we calculate what is the lepton pair radiation? It has
been shown \cite{gal91} that the thermal production rate for electron--positron
pairs is related to the imaginary part of the retarded photon self energy by

\begin{eqnarray}   \label{rate}
E_+E_- \frac{dR}{d^3p_+ d^3p_-} & = &
\frac {2e^2}{(2\pi)^6} \frac {1}{M^4}(p^\mu_+p^\nu_- + p^\nu_+p^\mu_-
-p_+\cdot p_- g^{\mu\nu})\ {\em Im} \Pi^{\rm R}_{(\gamma)\mu\nu}(k) \\
\nonumber
&  &\times
 \frac {1}{e^{E/T} -1}.
\end{eqnarray}
\noindent Here $p_+$ and $p_-$ are the positron and electron momenta,
$k^\mu = (E,\vec{k})$ is the virtual photon momentum,
$T$ is the temperature,
and we have set
the electron mass to zero (nonzero lepton mass is easy to include).
$R$ is the number of times per unit four-volume an $e^+e^-$ pair of invariant
mass $M$ is produced with the specified momentum configuration.
Note that the above equation is
perturbative in the electromagnetic interaction only; it is a completely
non--perturbative expression in the strong interaction.

Furthermore, we shall make use of the VMD model, which states
that the hadronic electromagnetic current operator is given by the {\em
current--field identity}
\begin{equation}
J_\mu = - \frac{e}{g_\rho} m^2_\rho \rho_\mu
-\frac{e}{g_\phi} m^2_\phi \phi_\mu
-\frac{e}{g_\omega} m^2_\omega \omega_\mu .\label{current}
\end{equation}

The above expression tells us how the electromagnetic radiation couples to
hadronic (in our case mesonic) matter: by first coupling to one of the vector
mesons with some coupling constant. In the above, we have kept
the $\rho, \omega$ and $\phi$ fields, but in some cases we shall tacitly
include also higher vector mesons by using phenomenological form factors
inspired by data.
We further need a model for how the mesons
interact among themselves. For this, we shall use a simple phenomenological
approach, inspired by the chiral properties of low energy QCD. Such classes of
phenomenological Lagrangians have been quite successful in the past in the
description of low energy hadronic physics \cite{mei88}. We are
explicitly interested in the interaction between the different possible
combinations of vector ($V$)  and pseudoscalar ($\varphi$) fields. For
reasons that will become clear shortly we restrict our discussion to the
following interaction Lagrangians \cite{mei88}:

\begin{equation}
{\cal L}^{\rm int}_{V V \varphi} = g_{V V \varphi} \
\epsilon_{\mu \nu \alpha \beta}\
\partial^\mu V^\nu \partial^\alpha V^\beta \varphi\  , \label{lag1}
\end{equation}
\noindent and
\begin{equation}
{\cal L}^{\rm int}_{V \varphi \varphi} = g_{V \varphi \varphi}\ V_\mu \varphi
\stackrel{\leftrightarrow}{\partial} \varphi\ . \label{lag2}
\end{equation}
\noindent In the above, the coupling constants are fitted for each field
combination, in a procedure we now describe. We have a model for how mesons
interact among themselves and how
they interact with the electromagnetic field. With this approach, let us study
a simple radiative process like the decay of a vector meson into a pseudoscalar
meson and a photon like {\em e.g.} $\omega\ \rightarrow \ \pi^0 \gamma$. In
this
model, the process goes via the $\omega \rho \pi$ vertex, owing to $G$ parity
conservation at the strong vertex,  and the $\rho^0$
couples to the photon in virtue of the current--field identity. This
corresponds
to the Feynman diagram of Fig. 1. The ratio of coupling constants from Eqs.
(\ref{lag1}) and (\ref{current}), in this
case $g_{\omega \rho \pi}/g_\rho$, is adjusted so that the correct experimental
radiative decay width \cite{pdt92} $\Gamma ( \omega\ \rightarrow\ \pi^0
\gamma)$ is
obtained. Our Lagrangians are then ``calibrated''
through all the following processes: $\rho\ \rightarrow\ \pi \gamma$,
${K^*}^\pm\ \rightarrow\ K^\pm \gamma$,
${K^*}^0({\bar {K^*}^0})\ \rightarrow\ K^0({\bar K^0}) \gamma$,
$\omega\ \rightarrow\ \pi^0 \gamma$, $\rho^0\ \rightarrow\ \eta \gamma$,
$\eta^\prime\ \rightarrow\ \rho^0 \gamma$, $\eta^\prime\ \rightarrow\ \omega
\gamma$, $\phi\ \rightarrow\ \eta \gamma$, $\phi\ \rightarrow \eta^\prime
\gamma$, $\phi\ \rightarrow \pi^0 \gamma$. One
realizes [{\em c.f.} Eqs. (\ref{current}, \ref{lag1}, \ref{lag2})] that via
this procedure, we can only fix the ratio of strong to
``electromagnetic'' (vector meson--photon)  couplings. However it is
this very combination we shall need
for our specific application.

We now integrate our model for interacting mesons with a dilepton radiation
calculation. If we keep a calculation of the photon self--energy at the
one--loop level an evaluation of its imaginary part, as instructed in Eq.
(\ref{rate}), will yield processes of the type V (PS) $\rightarrow$ PS (V)
$\gamma^*$, PS + PS $\rightarrow$ $\gamma^*$, V + PS $\rightarrow$ $\gamma^*$
and V + V $\rightarrow$ $\gamma^*$. Since such tree--level amplitudes can
be readily computed and that our general field--theoretic treatment for
dilepton emission has been shown to agree with relativistic kinetic
calculations (up to temperature--dependent effects in the form factors,
which have been shown to be small \cite{gal91}) we use the latter
approach. Finally note that the
two--body channels listed above will kinematically dominate the contributions
of the type V + PS $\rightarrow$ PS + $\gamma^*$, which we shall neglect. The
inclusion of such processes would correspond to evaluation of the photon
self--energy beyond the one--loop level.
The first attempt to investigate the role of processes with more
than two mesons involved has recently been made in  \cite{plpr}.
We will return to this point later.

The basic relativistic kinetic expression for the dilepton
production rate from a process $a\ +\ b\ \rightarrow\ e^+ e^-$
is well known and can be written down as
\begin{eqnarray}
   R_{a b \rightarrow e^+ e^-}\ & = &\ {\cal N}\
\int\ {{d^3 p_a}\over {2 E_a (2
\pi)^3}} \ {{d^3 p_b}\over {2 E_b (2\pi)^3}} \ {{d^3 p_+}\over {2 E_+
(2\pi)^3}} \ {{d^3 p_-}\over {2 E_- (2\pi)^3}}\ f_a\ f_b \\ \nonumber
 &  & \times\ \ \left| {\cal M} \right|^2 \   (2 \pi )^4  \ \ \delta^4
(p_a + p_b - p_+ - p_- )\ . \label{rate2}
\end{eqnarray}
Similarly, we may write a rate equation for the decay process
$a\ \rightarrow\ b\ +\ e^+e^-$:
\begin{eqnarray}
\label{rate3}
{{d R_{a\ \rightarrow\ b\ +\ e^+e^-}}\over{d M^2}}\ & = &\ {\cal N}\
\int {{d^3 p_a}\over {2 E_a (2
\pi)^3}} \ {{d^3 p_b}\over {2 E_b (2\pi)^3}} \ {{d^3 p_+}\over {2 E_+
(2\pi)^3}} \ {{d^3 p_-}\over {2 E_- (2\pi)^3}}\ f_a\ ( 1 + f_b ) \\ \nonumber
 &  & \times \left| {\cal M} \right|^2 \   (2 \pi )^4  \ \ \delta^4
(p_a - p_b - p_+ - p_- )\ \delta (M^2\ -\ (p_+\ +\ p_- )^2 ) .
\end{eqnarray}
\noindent In the above equations, $\cal N$ is an overall degeneracy factor
dependent upon the specific channel and the $f$'s are Bose--Einstein mean
occupation numbers.

These equations are not suitable for numerical evaluation because of the
delta functions. However, they can be cast, using standard methods of
simplifying phase integrals and the spherical symmetry
in momentum space, into an appropriate form.
The dilepton production rate for the
process $a\ +\ b\ \rightarrow\ e^+ e^-$ becomes
\begin{eqnarray}
\label{rate4}
    R_{a b \rightarrow e^+ e^-}\ & = &\ {{\cal N}\over{(2\pi)^4}}\
\int_{m_a}^\infty dE_a\ p_a f_a(E_a)
\int_{m_b}^\infty dE_b\ p_b f_b(E_b)   \\ \nonumber
 & &\times \int_{-1}^1 dx\ \sqrt{(s-s_+)(s-s_-)}\
\sigma_{a b \rightarrow e^+ e^-}(s),
\end{eqnarray}
\noindent where
$s\equiv M^2=m_a^2+m_b^2+2(E_aE_b-p_ap_bx)$,
$s_+=(m_a+m_b)^2$ and $s_-=(m_a-m_b)^2$.
The cross section
$\sigma_{a b \rightarrow e^+ e^-}(s)$,
is obtained by an evaluation of
an appropriate Feynman diagram, Fig. 2. The multiple integral in Eq.
(\ref{rate4}) is evaluated by Monte Carlo methods. While it is certainly
possible to do some of the integrations analytically, we chose the avenue of
keeping a relatively transparent integrand and we let the Monte Carlo approach
handle the numerical complexity. Moreover, this approach allows us to
evaluate any desirable differential dilepton production rate
easily (see, {\em e.g.},  \cite{pllvh} or  \cite{pljth}).

Similarly, the rate equation for the decay process $a\
\rightarrow\ b\ +\ e^+e^-$ is now:
\begin{eqnarray}
 \label{rate5}
{{d R_{a\ \rightarrow\ b\ +\ e^+e^-}}\over{d M^2}}\ & = &\
{{\cal N}m_a\over{(2\pi)^2}}\
{{d \Gamma_{a\ \rightarrow\ b\ +\ e^+e^-}}\over{d M^2}}\
\int_{m_a}^\infty dE_a\ p_af_a(E_a)
 \\ \nonumber
& & \times \int_{-1}^1 dx\ \left[ 1+f_b(E_b)\right],
\end{eqnarray}
where
$E_b=(E_a E_b^* + p_a p_b^* x )/m_a$,
$E_b^*=(m_a^2+m_b^2-M^2)/(2m_a)$ and $d \Gamma_{a\ \rightarrow\ b\ +\ e^+e^-} /
d M^2$ is the differential decay width into the appropriate channel.
In Eqs. (\ref{rate3}) and (\ref{rate5}) one  notices the Bose--Einstein
final state enhancement, an in--medium effect.

We also include the direct decay channels of the form $V~\rightarrow~e^+~e^-$.
As we will show, their contributions are non--negligible. This is especially
true in the case of $\rho~\rightarrow~e^+ e^-$.
One can show that for such decays
\begin{eqnarray}
{{d R_{\ V\ \rightarrow\ e^+e^-}}\over{d M^2}}\ & = &\
{{3 }\over{2 \pi^2}}\  {{\Gamma_{V\ \rightarrow\ e^+ e^-}}\over{\tilde{N}}}\
{{m_V^3}\over{M^2}}\ B ( M^2 )\  \int_{M}^{\infty} d E\ f ( E )\
\sqrt{E^2 - M^2}\ \ ,
\end{eqnarray}
where
\begin{eqnarray}
B ( M^2 ) & = &\ \beta\  {{\Gamma_{\rm tot}}\over{( M^2 -
m_V^2 )^2 + ( m_V \Gamma_{\rm tot} )^2}}\ \ .
\end{eqnarray}
The constant $\beta$ fixes the normalization of the Breit-Wigner probability
density function. Its value is not important here as it enters also the factor
\begin{eqnarray}
\tilde{N} & = &\ \int d M^2\ \left( {{m_V}\over{M}} \right)^3\ B ( M^2 )\ \ ,
\end{eqnarray}
which ensures the correct overall normalization based on the experimental value
of the partial decay width into the dielectron channel,
$\Gamma_{V~\rightarrow~e^+
e^-}$. The integral runs over the allowed mass range.

In the above equations, $m_V$ is the vector meson mass and  $\Gamma_{\rm tot}$
is its total decay width.   For the narrow resonances ($\omega, \phi$) the
latter
is taken constant but the $\rho^0$ width is given its proper mass dependence.

\section{Results}
\label{sec:results}

The decay channels considered have already been listed: they are the same
radiative
decay reactions  V (PS) $\rightarrow$ PS (V) + $\gamma$,
as used to fix the couplings constants of our Lagrangians, with the obvious
substitution: $\gamma$ $\rightarrow$ $\gamma^*$. The V + PS $\rightarrow$ $e^+
e^-$ amplitudes can all be obtained from the decay reaction amplitudes by
crossing
symmetry. We list the entrance channels anyway for completeness. They are:
$\omega\ \pi^0$, $\rho\ \pi$,   $\phi\ \pi^0$,
   $\omega\ \eta$, $\phi\ \eta$, $\rho^0\ \eta$, $\omega\ \eta^\prime$,
  $\phi\ \eta^\prime$, $\rho\ \eta^\prime$,
${\bar K^*}K$ and $K^*{\bar K}$.
For each of the PS + PS and V + V reaction, we follow the following
approach:
their ``bare'' amplitude is calculated, squared, and finally
multiplied by
a form factor obtainable from experimental data
on $e^+e^-$ annihilation.

The topic of form factors deserves here a short discussion. Of course no
information on time--like form factors is available through the analysis of
meson radiative decays into real photons. With respect to this issue, we have
followed a simple prescription.
The time--like electromagnetic form factor of charged pion is
experimentally very well known \cite{exffpi} and some experimental
information exists also about those of both charged and neutral kaons
 \cite{exffK}. In our calculations of $\pi^+\pi^-$, $K^+K^-$, and
$K^0{\bar K^0}$ annihilation rates
we have used a recent parametrization \cite{dub91} of these quantities.
The vector mesons annihilation channels have been
given the same form factors as their corresponding (by strangeness
and isospin) pseudoscalar counterparts.
In the case of decays and V + PS reactions, whenever
the $G$ parity and isospin conservation laws allowed a coupling only to
the $\rho ^0$ and its recurrences, the charged pion electromagnetic form
factor \cite{dub91} was used. In the other cases, we have stuck with a form
factor equivalent to a simple pole corresponding to the lightest permitted
vector meson.
Our way of normalizing coupling constants by means of the radiative decay
widths leads us to a belief that this conservative choice of form factors
does not introduce too much uncertainty.
We made only one exception from the simple rules sketched above. In the
case of the reaction
$\rho + \pi \rightarrow e^+e^-$ the rules would lead to a simple
$\omega$--pole. It would be a rather bad approximation because the threshold
of this reaction lies below the position of the $\phi$--resonance, which
thus becomes extremely important. We take therefore a two--pole formula
with the relative weight between the $\omega$ and $\phi$ contributions
same as in the kaon isoscalar form factor
$F_S=(F_{K^+}+F_{K^-})/2$\cite{dub91}.

We have performed our thermal hadronic calculations at three
temperatures: 100, 150 and 200 MeV. We feel that those reflect a
range of energies that is somewhat reasonable, by current theoretical
standards.

The results for V (PS) $\rightarrow$ PS (V) $e^+ e^-$ at a temperature
of 150 MeV are
shown on Fig. 3. Not all the decays are shown, but only
the dominant ones. Coupling constants arguments aside, the largest
contributions will come from the radiative channels where a heavy meson
decays into a light one and a lepton pair. This is precisely what is
observed on Fig. 3. The largest contribution up to invariant mass
$\approx$ 0.65 GeV is from $\omega\ \rightarrow\ \pi^0 \ e^+ e^-$. Over this
range, $\rho\  \rightarrow\ \pi\ e^+ e^-$ represents the next--to--leading
contribution and the other decays are at least an order of magnitude
lower. Above 0.65 GeV invariant mass, the only decay with phase
space left is $\phi\ \rightarrow\ \pi^0\ e^+ e^-$. Note that the widths
for the radiative decays of the $\omega$ and $\rho^0$ are comparable and
are two orders of magnitude larger than that for $\phi\ \rightarrow\
\pi^0 \gamma$ \cite{pdt92}.
Dalitz decay ({\em e.g.} $\eta\ \rightarrow\ \gamma
e^+ e^-$) is of higher order in $\alpha$ and can thus be neglected.
However, this argument alone is not totally convincing as one could
imagine that the $\eta$ could be massively produced at such
temperatures.
We have therefore performed a calculation of the contribution from
eta Dalitz decay to thermal electron pair yield, using the VMD
prescription for $d\Gamma/dM^2$ \cite{bud79} with updated coupling
constants. We have
found it in fact
to be orders of magnitude smaller than the channels discussed above.

For the pseudoscalar--pseudoscalar reactions, on Fig. 4 we display a
plot of all
the contributions, again at $T = 150$ MeV. The different contributions add
up to a signal in which the only apparent structures are associated with
the $\rho ( 770 )$ and the $\phi$, with a slight shoulder at the
$\rho ( 2150 )$. The
peak in the pion form factor at the $\rho ( 1700 )$ is washed out by the
kaon contributions.

We show the V + V contributions on Fig. 5. Above
threshold, the sum of these processes outshine the PS + PS ones by
roughly an order of magnitude. The structure at $M = 2.15$ GeV owes to the
corresponding excitation of the $\rho$.

The V + PS reactions are quite numerous, we show the brighter dilepton
sources on Fig. 6, again for $T = 150$ MeV. The dominant channels are
$\omega\ + \ \pi^0$, $\rho\ +\ \pi$ and $\rho^0\ +\ \eta$. The kaon
channels are not shown but are roughly the size of the $\pi\ +\ \rho$
contribution. The strongest signal is from $\omega\ + \ \pi^0$, over the
entire invariant mass range considered here. Recall from our discussion
of the decays that the radiative decay widths of the $\rho$ and the
$\omega$ are quite large.

Finally, the total rate corresponding to the sum of all processes discussed so
far is shown on Fig. \ref{totrate150}, along with a curve representing the
$\pi^+\pi^-$ contribution only. We also show the net direct decay contribution,
summing
$\rho\ \rightarrow\ e^+ e^-$, $\omega\ \rightarrow\ e^+ e^-$ and $\phi\
\rightarrow\ e^+ e^-$. The radiation from these channels turns out to be quite
important. The signal from the decay reaction $\rho~\rightarrow~e^+ e^-$
closely resembles the pion annihilation spectrum, which in retrospect is
quite reasonable.

In all cases (decays, PS + PS, V + V, V + PS)
our findings at $T = 100$ and 200 MeV are qualitatively similar, with a
global shift in the rate. For these temperatures we therefore present
only the total rates (see Fig. \ref{totrates}).
\section{Discussion}
\label{sec:conclusion}

Up to now, thermal calculations of the variety discussed in this paper have
rarely gone beyond a pure pion gas approximation, usually concentrating on the
annihilation channel \cite{kaj86}. The contribution from thermal
meson decays has been considered previously \cite{koc92}. To our knowledge it
is
the first time that extensive mesonic reactions have been included, together
with direct decays.

Comparing the individual contributions from different processes (Figs. 3--6)
to the total dilepton rate (Fig. \ref{totrate150}) one sees that the
dilepton invariant
mass spectrum naturally divides in several parts. At low masses, the decay
channels clearly dominate the entire spectrum. The crossover to the pion--pion
annihilation and direct decay signal occurs just above 0.5 GeV (at the
lower temperature,
$T = 100$ MeV, this crossing point is shifted closer to the two--pion
threshold). Already at $M\ \approx$ 1 GeV, the
total rate dominates over the pion gas
approximation result by an approximate factor of 3. At $M$ = 1.5 GeV, those
rates
differ by a little more than an order of magnitude. The difference increases
with larger invariant masses. One also sees that the net rate at the vector
meson positions is also larger than in the straight $\pi^+ - \pi^-$ scenario,
owing principally to direct decays and also form factor effects. Probably the
most striking conclusion of our work
is that the ``usual'' pion results for lepton pair production calculation
holds rather poorly over all regions of invariant masses considered in this
work. This statement is true for all temperatures studied here.

Thus, the rate for $M\ \agt$ 1 GeV is approximately
one order of magnitude larger in our calculation than in ``conventional'' meson
background calculations. This enhancement is also present in the
momentum structure of the lepton signal: Fig. \ref{momspect} is a plot
of $E d^3 R/d^3 p$ for lepton pair invariant masses between 1.1 and
3 GeV.
These findings should have important implications in connection
with the plasma signal identification. The conventional window for
             thermal lepton pairs of plasma origin  is
$m_\phi\ <\ M\ < m_{J/\psi}$
\cite{rus92}, precisely the range discussed here. An observation of
a signal from an exotic source can only be claimed if all other sources are
under control. Here these would be identified with the Drell--Yan mechanism,
open charm decay \cite{shor} and the thermal background we have considered in
this work. However, before any more quantitative statements can be made, it is
imperative to complement our calculations with a dynamical model of some sort,
in order to make contact with genuine observables. Work in this direction is in
progress.

It is of interest to compare the rates obtained with other similar
calculations. Some recent interest has been devoted to the emission of lepton
pairs from pionic bremsstrahlung processes \cite{brem92}. It was concluded that
the radiation from the external pion lines in pion--pion collisions would be a
dominant contribution to the low mass lepton spectrum. Comparing with pion
bremsstrahlung calculations at $T=150$ MeV, we realize that the low mass signal
is the same magnitude as the net meson decay contribution. Correcting the
pion--pion bremsstrahlung rate for the Landau--Pomeranchuk effect \cite{lp92}
will cut this pion signal by some factor. This factor is only $\approx 2$ for
low invariant masses and $T=150$ MeV \cite{cley93}. This correction
also goes down as invariant mass grows. This will then leave the pion
bremsstrahlung to compete with the decay channels contribution, up
to the two--pion annihilation threshold.

In this inquiry, we have pursued the same goals as a
similar photon production calculation \cite{kap91}. Our main aim has
been to identify the most important dilepton production processes
which operate in a hadron gas. We considered only the decays and reactions
with the minimal possible number of hadrons: one in the decay final states,
none in the final states of two-initial-hadron reactions. These processes
are believed, on the basis of the phase-space
and order-of-interaction arguments, to be dominant here. The reactions
of this kind ($2\ \rightarrow\ 0$ hadrons) do not operate in real photon
production due to restrictions from energy-momentum conservation.
However, the dominant reactions for photon
production $a+b\rightarrow c+\gamma$ can produce virtual photons as well.
It is clear that they would populate preferably the low-mass region.
Even there they would be probably negligible, as it was shown
for the case of $\pi +\pi\ \rightarrow\ \pi +$~dilepton in \cite{plpr}.
But one cannot exclude surprises. The latter process amplifies, together
with the three pion annihilation channel, the omega peak in dilepton
spectrum. This may in turn serve as a signature of a hadron gas
creation \cite{plpr}.
It has also been pointed out that the $A_1$ meson could have a
significant influence on the real photon yield \cite{xio92}, through
the process $\pi\ \rho\ \rightarrow\ A_1\ \rightarrow\ \pi\ \gamma$.
This conjecture has been carefully analyzed in a recent paper \cite{song93}.
The reflection in the thermal dilepton sector is certainly worth studying
as well. Three body initial state processes $a+b+c\rightarrow \ e^+e^-$
may also contribute significantly in the high invariant mass region
\cite{plpr}.
We intend to study all these points in detail in future work.

\acknowledgements

We  would like to acknowledge the warm hospitality of the Theoretical
Physics Institute of the University of Minnesota, where this work was
started. This work
was supported in part by the Natural Sciences and Engineering Research
Council of Canada, by the FCAR fund of the Qu\'ebec Government and by a
NATO Collaborative Research grant.
The stay of P.L. at the University of Minnesota was supported by the U.S.
Department of Energy under Contract No. DOE/DE-FG02-87ER-40328;
travel expenses were borne by the grant M\v{S}M\v{S}~SR~01/35.


\begin{figure}
\caption{The Feynman diagram for the radiative decay
$\omega\rightarrow\pi^0\gamma$ in the model described in the text.}
\end{figure}

\begin{figure}
\caption{A ``generic" two--body amplitude with lepton pairs in the
final state. The different $\{a,b\}$ combinations we consider are
enumerated in the text. The vector meson V is chosen through isospin
and $G$ parity arguments.}
\end{figure}

\begin{figure}
\caption{Differential rate for lepton pair production via vector or
pseudoscalar meson decay. The dashed line represents the contribution
from $\omega\ \rightarrow\ \pi^0 e^+ e^-$, the dashed--dotted line is
the rate from $\rho\ \rightarrow\ \pi e^+ e^-$. The dotted line is the
process $\phi\ \rightarrow\ \pi^0 e^+e^-$. The structure in the latter
channel is due to the $\rho ( 770 )$.  The solid line is the sum of all
the decay processes, including those not listed in this caption but
enumerated in the
main text.}
\end{figure}

\begin{figure}
\caption{Rate from PS + PS type reactions. The dashed line is the rate for
the pion annihilation process. The dotted curve represents the
contribution from $K^+\ +\ K^-$. The dashed--dotted line is the rate
from $K^0\ {\bar K}^0$ annihilation. The solid line is the sum of the PS + PS
processes.}
\end{figure}

\begin{figure}
\caption{Rate from V + V type reactions. The dashed curve is the
$\rho^+\ + \rho^-$ contribution. The dashed--dotted and dotted
curves represent charged and neutral $K^*$ annihilation, respectively. The
solid curve is the sum of the V + V contributions.}
\end{figure}

\begin{figure}
\caption{Rate from V + PS type reactions. The dashed curve is the rate
from $\omega\ +\ \pi^0$. The dashed--dotted curve is the contribution
from $\rho\ +\ \pi$ and the dotted curve is the rate from $\rho^0\ +\
\eta$. Again, the solid line is a sum of all V + PS processes, as enumerated in
the text.}
\end{figure}

\begin{figure}
\caption{The solid line is the total rate
at $T=150$ MeV from all processes discussed
in the text. The dashed line is the pion--pion annihilation contribution
only. The short--dashed curve represents the contribution from direct vector
meson decays.}
\label{totrate150}
\end{figure}

\begin{figure}
\caption{Same caption as Fig. 7 but for the
temperatures $T=100$ MeV
(lower curves) and $T=200$ MeV (higher curves).}
\label{totrates}
\end{figure}

\begin{figure}
\caption{The lepton pair momentum spectrum, $E d^3 R/d^3 p$ for
lepton pair invariant masses between 1.1 and 3 GeV. This lower bound is chosen
so as to exclude the $\phi$ peak.  The full curve
represents the contribution from all processes described in this work.
The dashed curve is the pion--pion annihilation contribution only.}
\label{momspect}
\end{figure}

\end{document}